\documentclass[twocolumn]{aastex62}
\pdfoutput=1 
\usepackage{amsmath,amstext}
\usepackage{apjfonts} 
\usepackage[figure,figure*]{hypcap}
\usepackage{graphicx}	
\usepackage{amssymb}	
\usepackage{booktabs}
\usepackage{chemfig}
\usepackage[percent]{overpic}

\received{XXX}
\revised{XXX}
\accepted{XXX}

\begin{document}
\title{HyDRA-H: Simultaneous Hybrid Retrieval of Exoplanetary Emission Spectra}

\correspondingauthor{Siddharth Gandhi, Nikku Madhusudhan}
\email{Siddharth.Gandhi@warwick.ac.uk, nmadhu@ast.cam.ac.uk}


\author{Siddharth Gandhi}
\affiliation{Institute of Astronomy, University of Cambridge, Madingley Road, Cambridge, CB3 0HA, UK}
\affiliation{Department of Physics, University of Warwick, Coventry CV4 7AL, UK}
\author{Nikku Madhusudhan}
\affiliation{Institute of Astronomy, University of Cambridge, Madingley Road, Cambridge, CB3 0HA, UK}
\author{George Hawker}
\affiliation{Institute of Astronomy, University of Cambridge, Madingley Road, Cambridge, CB3 0HA, UK}
\author{Anjali Piette}
\affiliation{Institute of Astronomy, University of Cambridge, Madingley Road, Cambridge, CB3 0HA, UK}
\begin{abstract}
High-resolution Doppler spectroscopy has been used to detect several chemical species in exoplanetary atmospheres. Such detections have traditionally relied on cross correlation of observed spectra against spectral model templates, an approach that is successful for detecting chemical species but not optimised for constraining abundances. Recent work has explored ways to perform atmospheric retrievals on high-resolution spectra (HRS) and combine them with retrievals routinely performed for low-resolution spectra (LRS) by developing a mapping from the cross correlation function to a likelihood metric. We build upon previous studies and report HyDRA-H, a hybrid retrieval code for simultaneous analysis of low- and high- resolution thermal emission spectra of exoplanets in a fully Bayesian approach. We demonstrate HyDRA-H on the hot Jupiter HD~209458b as a case study. We validate our HRS retrieval capability by confirming previous results and report a simultaneous hybrid retrieval using both HRS and LRS data. The LRS data span the HST WFC3 (1.1-1.7 $\mu$m) and Spitzer photometry (IRAC 3.6-8$\mu$m) bands, while the HRS data were obtained with CRIRES on VLT at 2.3 $\mu$m. The constraints on the composition and temperature profiles for the hybrid retrieval are more stringent than retrievals with either LRS or HRS datasets individually. We retrieve abundances of $\log(\mathrm{H_2O)} = -4.11^{+0.91}_{-0.30}$ and $\log(\mathrm{CO}) = {-2.16}^{+0.99}_{-0.47}$, and C/O = 0.99$^{+0.01}_{-0.02}$, consistent with previous works. We constrain the photospheric temperature to be $1498^{+216}_{-57}$~K, consistent with the equilibrium temperature. Our results demonstrate the significant advantages of hybrid retrievals by combining strengths of both HRS and LRS observations which probe complementary aspects of exoplanetary atmospheres. 
\end{abstract}

\section{Introduction}
\label{introduction}
The characterisation of exoplanet atmospheres has experienced rapid growth in recent years as an increasing number of spectra are becoming available with ever-improving data quality. Constraints on the chemical abundances, C/O ratios and thermal profiles of several exoplanets have been obtained using a wide variety of observations from both space-based and ground-based telescopes, and in both the optical and infrared wavelength ranges. Low-resolution spectra from both space and the ground have allowed for unprecedented chemical characterisation thanks to the photometric precision achieved. In particular, spectra from the Hubble Space Telescope (HST) have led to constraints on the chemical and thermal properties of several exoplanets \citep[e.g.][]{kreidberg2014,madhu2014,sing2016,sheppard2017,evans2017}. 

A variety of techniques have been used to extract atmospheric properties from low-resolution spectra (LRS). In particular, atmospheric retrievals provide a data-driven approach to inferring properties such as chemical compositions and thermal profiles \citep{madhu2018}. This method involves coupling an atmospheric model to a parameter-estimation algorithm in order to fit observed spectra and derive statistical constraints on the model parameters. A variety of both models and Bayesian estimators have been used in the literature \citep[e.g.][]{madhu2009,madhu2010,line2013,line2016,barstow2017,lavie2017,evans2018,benneke2019}. The model parameters span a wide range of chemical compositions and temperature profiles for a given observing geometry, e.g. transmission or emission spectra. Other model considerations include treatment of clouds/hazes \citep{line2016_clouds,barstow2017,macdonald2017}, radiative disequilibrium \citep{gandhi2018}, stellar heterogeneities \citep{pinhas2018}, and multi-dimensional effects \citep{feng2016, blecic2017}.
 
Emission spectra of transiting planets observed at secondary eclipse provide a unique window into the dayside of their atmospheres. In particular, such spectra probe a wide range of pressures and temperatures, allowing the thermal profile of the atmosphere to be constrained. Since the thermal emission of exoplanets peaks in the infrared, observations to date have largely been restricted to this spectral range. In particular, near-infrared spectra and photometry from HST and Spitzer have resulted in constraints on multiple chemical abundances as well as the inference of thermal inversions in several exoplanets \citep[e.g.][]{Haynes2015,evans2017,sheppard2017,kreidberg2018}. However, most compositional constraints have been obtained for H$_2$O given the spectral range of these observations. Retrievals of emission spectroscopy have not only been performed on transiting planets, but also on directly-imaged planets and field brown dwarfs, providing insight into the properties of a wide range of sub-stellar objects \citep[e.g.][]{line2017,burningham2017,lavie2017}.

On another front, high-resolution Doppler spectroscopy (HRS) has enabled the detection of several chemical species in exoplanet atmospheres \citep[e.g.][]{snellen2010, brogi2012, birkby2017,nugroho2017}. These ground-based observations resolve individual transition lines from the planetary spectrum, which are distinguished from the stellar spectrum thanks to the Doppler motion of the planet as it orbits the host star \citep[see e.g. review by][]{birkby2018}. Thanks to stable high-resolution spectrographs on telescopes such as the Very Large Telescope (VLT), Keck and Subaru, HRS has begun to play a significant role in the characterisation of exoplanet atmospheres \citep[e.g.][]{brogi2012,nugroho2017}. Molecules such as H$_2$O and CO have been detected in multiple hot Jupiters using this method \citep{snellen2010, brogi2012, birkby2013, rodler2013, birkby2017}. One of the key strengths of high-resolution Doppler spectroscopy is its ability to detect trace species in exoplanet atmospheres. For example, TiO has been detected in WASP-33b \citep{nugroho2017}, HCN has been seen in multiple planets \citep{hawker2018, cabot2019}, and various atomic and ionic species have also been observed using high-resolution observations \citep{hoeijmakers2018}. In the next decade, this new era of atmospheric observations is expected to flourish thanks to upcoming facilities such as the Extremely Large Telescope (ELT) and the Giant Magellan Telescope (GMT), which will allow an increasing range of chemical species to be probed for ever-smaller planets, potentially allowing for the characterisation of rocky exoplanets \citep{rodler2014,snellen2015}.

More recently, retrievals have begun to be adapted for high-resolution spectra as well. HRS observations have traditionally been analysed by cross-correlation with model spectra, and detections of chemical species were made by exploring spectra over a grid of atmospheric parameters \citep[e.g.][]{brogi2012,birkby2017,nugroho2017}. However, a grid-search approach does not offer a robust statistical exploration of the parameter space, thereby precluding statistical estimates of the model parameters given the data. This is in contrast to the retrievals performed on low-resolution spectra, as discussed above. \citet{brogi2017} first performed a joint analysis of high-resolution and low-resolution observations for HD~209458b. \citet{brogi2019} went on to suggest a cross-correlation to log-likelihood mapping and performed the first, fully Bayesian retrieval on HRS observations. This method not only allows the presence of chemical species to be inferred, but also provides some constraints on their abundances and the thermal profile of the atmosphere using HRS observations.

Combined retrievals with low- and high-resolution spectra offer new promise for atmospheric characterisation. Since LRS and HRS are sensitive to different atmospheric properties, analysing both types of observations in parallel allows more robust constraints to be made on the properties of exoplanet atmospheres. For example, LRS observations are more sensitive to absolute abundances than HRS, but are affected by degeneracies between the overlapping opacities of different species in the observed spectral bands. However, HRS is sensitive to the unique fingerprints of each species and could potentially be used to break these degeneracies. Moreover, HRS observations probe regions higher up in the atmosphere (i.e. lower pressures) compared to low-resolution spectra. \citet{brogi2019} developed a fully Bayesian framework for combining LRS and HRS observations in atmospheric retrievals of thermal emission spectra of exoplanets. In this work, we demonstrate this framework on combined LRS and HRS observations of HD~209458b.

We report HyDRA-H, a new hybrid retrieval code building upon our existing HyDRA retrieval framework for low-resolution spectra \citep{gandhi2018}, and adopting the high-resolution retrieval framework of \citet{brogi2019}. We use this method to perform a fully Bayesian, simultaneous hybrid retrieval on observations of HD~209458b. The low-resolution spectrum we use is from the Hubble Space Telescope's Wide Field Camera 3 (HST/WFC3) in the 1.1-1.7 $\mu$m range and Spitzer's Infrared Array Camera (IRAC) at 3.6 $\mu$m, 4.5 $\mu$m, 5.8 $\mu$m and 8.0 $\mu$m. For the high-resolution spectrum we use archival observations at 2.3~$\mu$m obtained using the CRIRES instrument on the VLT \citep{snellen2011_survey}. With the results from this hybrid retrieval, we explore the chemical detections, abundances and $P$-$T$ profile of the dayside atmosphere of HD~209458b. 

In what follows, we discuss the retrieval method and its architecture in section \ref{sec:methods}. In section \ref{sec:hiresobs}, we outline the high-resolution data used for this work as well as its detrending and analysis. We validate the method against published results in section \ref{sec:validation}. Using these methods, we perform a hybrid retrieval on the combined HRS and LRS observations of HD~209458b, presenting our results in section \ref{sec:results}. Our conclusions are discussed in section \ref{sec:conclusion}.

\section{Methods}\label{sec:methods}

In this work we develop a hybrid retrieval code, HyDRA-H, capable of simultaneously retrieving low-resolution and high-resolution emission spectra of transiting exoplanets. In the past, retrievals have typically been designed for low-resolution spectra only, while high-resolution spectra were analysed separately using cross-correlation methods. Recently, \citet{brogi2017} presented the first joint analysis of low- and high-resolution spectra on the dayside emission spectra of HD~209458b. This method involved performing a retrieval on low-resolution data and using the resulting posterior probability distributions as priors in the analysis of the high-resolution data. Following this, \citet{brogi2019} made a further step by developing a joint retrieval framework for low- and high-resolution spectra to be analysed in parallel. To do this, they developed a metric to calculate the likelihood of model spectra based on the cross-correlation of the model with the high-resolution data. They demonstrate the capabilities of this method on simulated data, showing that the low- and high-resolution spectra contribute complimentary constraints on the atmospheric model parameters. Furthermore, this method has the advantage of allowing both types of spectra to freely explore the full parameter space. 

\begin{figure}
    \centering
    \includegraphics[width=0.47\textwidth]{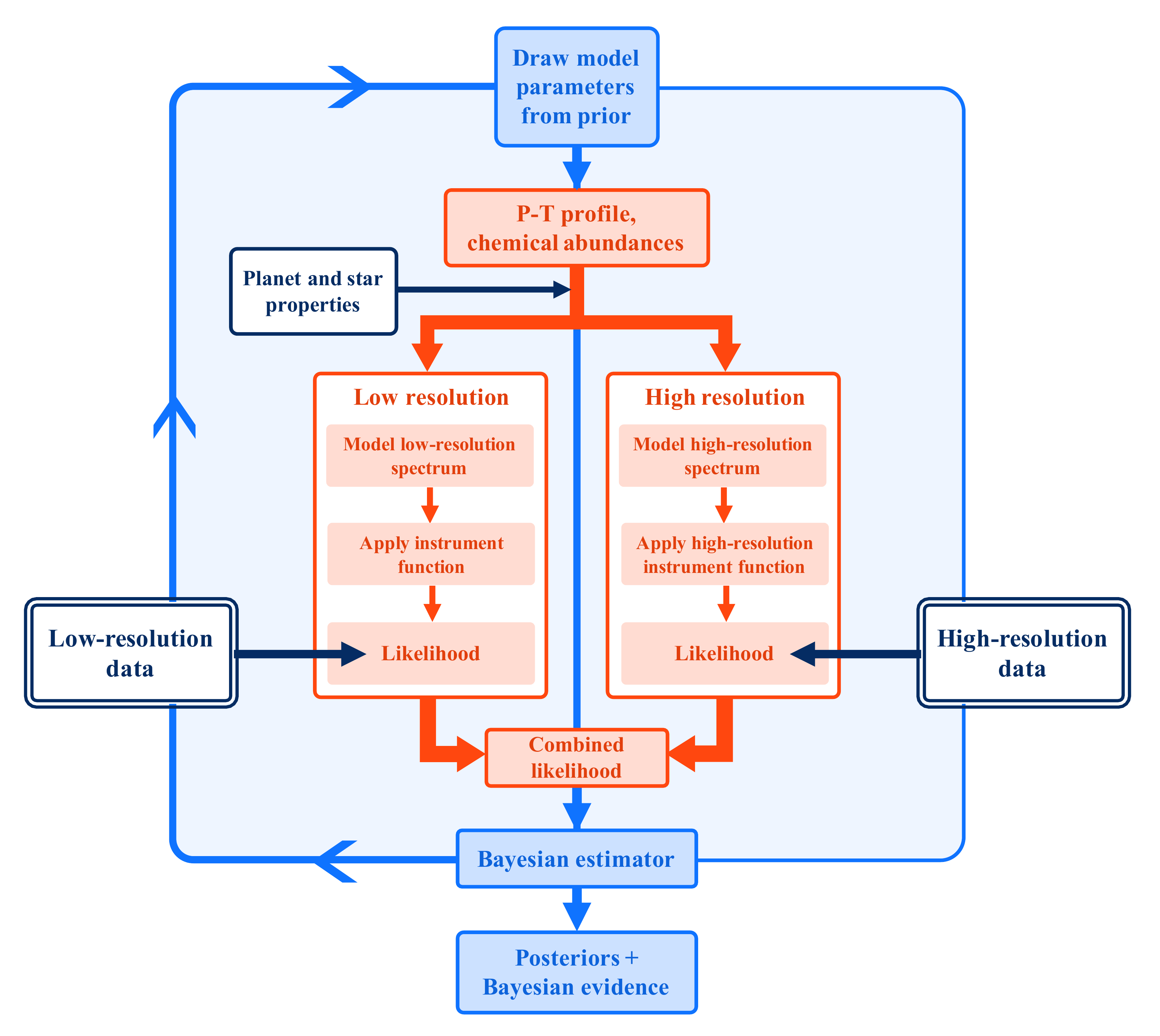}
    \caption{The hybrid modelling and retrieval framework. Model parameters are drawn from the priors listed in Table 1, and used to generate two types of model spectra: low-resolution for comparison with low-resolution data, and high-resolution for comparison to high-resolution Doppler spectroscopy.}
    \label{fig:flowchart}
\end{figure}

Here, we adapt the HyDRA retrieval framework \citep{gandhi2018} to analyse high-resolution data as well as low-resolution spectra, using the methods from \citet{brogi2019} to treat the high-resolution models and data. We then apply this hybrid retrieval method, HyDRA-H, to demonstrate the first simultaneous retrieval on low- and high-resolution spectra of HD~209458b. The architecture of HyDRA-H is shown in Figure \ref{fig:flowchart}. In this section, we first describe the retrieval framework and the calculation of model spectra in section \ref{sec:architecture}. We then outline how a combined LRS-HRS likelihood is calculated for each model spectrum in section \ref{sec:likelihoodcombo}.

\begin{table}
    \centering
    \begin{tabular}{c|c|c}
        \textbf{Parameter} & \textbf{Prior} & \textbf{Prior Range} \\
        \hline
        $X_{\mathrm{H}_2\mathrm{O}}$, $X_{\mathrm{CO}}$, $X_{\mathrm{CO}_2}$, $X_{\mathrm{HCN}}$ & log-uniform & $10^{-15}-1$ \\
        $T_{100\mathrm{mb}}$/$\mathrm{K}$ & linear & 300 - 2800 \\ 
        $\alpha_1$/$\mathrm{K}^{-1/2}$ & linear & 0 - 1 \\
        $\alpha_2$/$\mathrm{K}^{-1/2}$ & linear & 0 - 1 \\
        $P_1$/bar & log-uniform & $10^{-6}-10^{2}$\\
        $P_2$/bar & log-uniform & $10^{-6}-10^{2}$\\
        $P_3$/bar & log-uniform & $10^{-2}-10^{2}$\\
        d$K_\mathrm{p}$/kms$^{-1}$ & linear & -50 - 50 \\
        d$V_{\mathrm{sys}}$/kms$^{-1}$ & linear & -50 - 50 \\
        log($a$) & linear & -2 - 2
    \end{tabular}
    \caption{Priors and prior ranges for the model parameters: molecular mixing fractions ($X_{\mathrm{H}_2\mathrm{O}}$, $X_{\mathrm{CO}}$, $X_{\mathrm{CO}_2}$, $X_{\mathrm{HCN}}$), $P$-$T$ profile parameters ($T_{100\mathrm{mb}}$, $\alpha_1$, $\alpha_2$, $P_1$, $P_2$, $P_3$) and Doppler spectroscopy parameters (d$K_\mathrm{p}$, d$V_{\mathrm{sys}}$, log($a$)).}
    \label{tab:params}
\end{table}

\begin{figure*}
\centering
	\includegraphics[width=\textwidth,trim={0cm 0 0cm 0},clip]{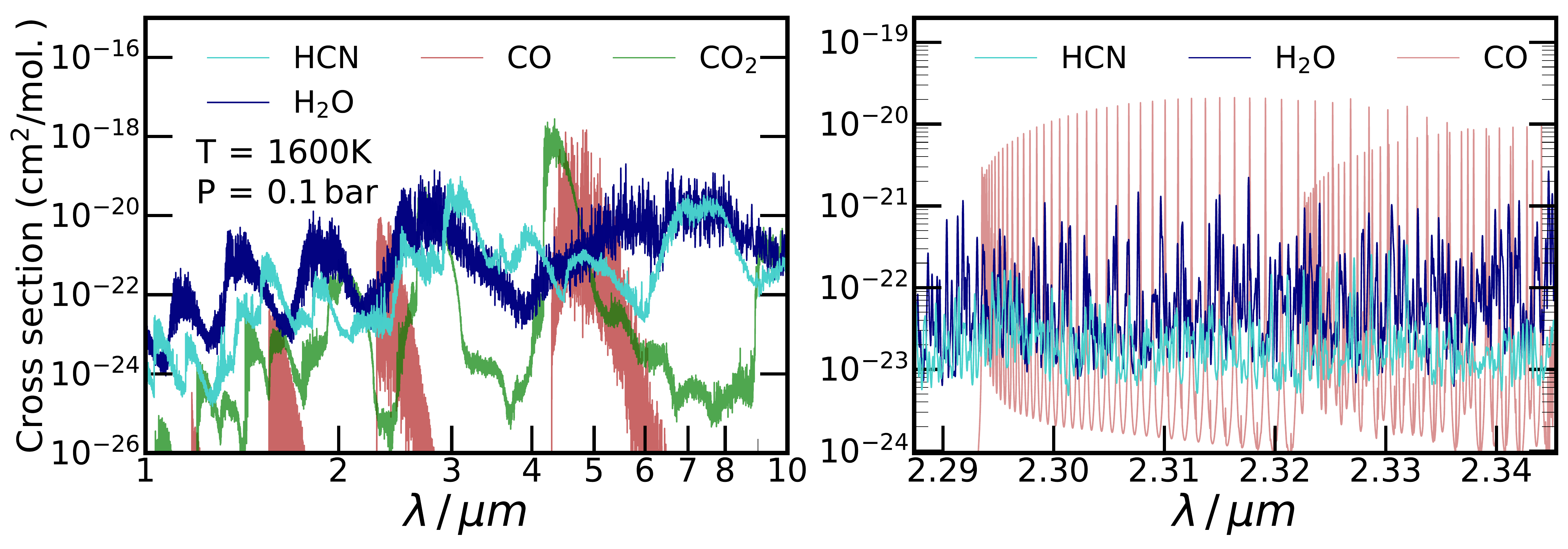}
    \caption{Molecular cross sections for the four species retrieved in the hybrid retrieval. In the left panel we show the cross sections between 1-10~$\mu$m for the HST WFC3 and Spitzer observations and the 2.29-2.35~$\mu$m CRIRES range is shown on the right.}     
\label{fig:cs}
\end{figure*}
\subsection{Retrieval Architecture}
\label{sec:architecture}

The central element of the retrieval framework is the calculation of model spectra and their comparison to low-/high-resolution data. In this work, we calculate model spectra using an adaptation of the HyDRA retrieval framework \citep{gandhi2018}. Each atmospheric model in the retrieval is defined by parameters describing its chemical abundances and $P$-$T$ profile. The model computes radiative transfer in a plane-parallel atmosphere, assuming hydrostatic equilibrium and an ideal gas equation of state. The methodology used to compute radiative transfer is described in \citet{gandhi2018}. The chemical species we include in this work are H$_2$O, CO, CO$_2$ and HCN, which are known to have strong opacity in hot Jupiter atmospheres, and their mixing fractions are assumed to be constant with depth. The molecular cross sections for these species are calculated as in \citet{gandhi2017} using line lists from the HITEMP database for H$_2$O, CO and CO$_2$ \citep{rothman2010} and the ExoMol database for HCN \citep{harris2006, barber2014, tennyson2016}. Figure \ref{fig:cs} shows the cross sections for the low- and high-resolution spectral ranges. CO$_2$ has a very weak cross section in the 2.29-2.35~$\mu$m spectral range compared to the other species and thus we do not include this species in our HRS calculations. The CO$_2$ can however show strong degeneracies with CO in the Spitzer 4.5~$\mu$m band due to their similar strong cross sections. In addition to the molecular absorption, all of our spectral models also include collisionally induced absorption (CIA) from H$_2$-H$_2$ and H$_2$-He interactions \citep{richard2012}. We parameterise the $P$-$T$ profile analytically using the method of \citet{madhu2009}, which requires 6 parameters (Table \ref{tab:params}). Once a model spectrum is generated, the relevant instrument function is applied before it is compared to the data and a likelihood is calculated. The Bayesian inference and parameter estimation is performed using PyMultiNest \citep{feroz2008,feroz2009,feroz2013,buchner2014}, a Nested Sampling Bayesian parameter estimation algorithm \citep{skilling2006}. 

The HyDRA-H retrieval framework is shown in Figure \ref{fig:flowchart}. This differs from the HyDRA retrieval framework as a high-resolution spectrum is computed as well as a low-resolution spectrum. In particular, this method takes advantage of the fact that low- and high-resolution spectra are sensitive to different atmospheric properties to optimise the spectral calculations. Once a set of atmospheric parameters is drawn by the Bayesian estimator, the low- and high-resolution model spectra are computed separately, and each compared to the low- and high-resolution data, respectively. Since low-resolution emission spectra are sensitive to the absolute planet-star flux ratio, the low-resolution model is computed using the more accurate double-ray quadrature scheme for radiative transfer, as described by \citet{gandhi2018}. Due to the self-calibration of high-resolution data (described in section \ref{sec:hiresobs}), sensitivity to absolute flux levels is reduced. We therefore use single-ray radiative transfer to compute the high-resolution models, for computational efficiency. In contrast, the high-resolution observations resolve individual molecular lines and are very sensitive to the presence of individual molecules as well as their relative abundances. The high-resolution model spectra are therefore calculated at a wavenumber spacing of 0.01cm$^{-1}$ ($R\sim 5\times 10^{5}$ at 2~$\mu$m). The low-resolution models are calculated at a slightly lower spectral resolution \citep{gandhi2017} using cross sections with a wavenumber spacing of 0.1cm$^{-1}$.

In addition to the chemical and thermal model parameters, three extra parameters are needed to compare the high-resolution models to high-resolution data. The parameters d$K_\mathrm{p}$ and d$V_{\mathrm{sys}}$ are perturbations to the known planetary radial velocity semi-amplitude and the systemic velocity, respectively. Including these parameters prevents the retrieval from being biased by uncertainties in the measurements of $K_\mathrm{p}$ and $V_{\mathrm{sys}}$. Their retrieved values are expected to be consistent with zero, which indicates that the planetary signal is detected at the expected location in $K_\mathrm{p}$-$V_{\mathrm{sys}}$ space. We also include a scaling parameter, log($a$), which is discussed further in section \ref{sec:HRCCloglikelihood}. The full list of model parameters used and their priors are shown in Table \ref{tab:params}.

\subsection{Observations}

Here we discuss the LRS and HRS observations for the retrievals conducted in this work. We obtain dayside emission spectra for HD~209458b at low and high spectral resolution covering a wide spectral range. We retrieve the LRS-only, HRS-only and the combined data in separate retrievals in this work, which is discussed in detail in sections \ref{sec:validation} and \ref{sec:results}. The LRS data consists of HST WFC3 and Spitzer observations and the HRS data is obtained from archival CRIRES observations \citep{snellen2011_survey}. The low resolution HST observations have been obtained from \citet{line2016} and span the 1.1-1.7~$\mu$m range. The Spitzer photometric observations cover the IRAC 1-4 bands in the $\sim$3-10~$\mu$m range \citep{diamondlowe2014}.

The high-resolution K-band observations of HD~209458b have been obtained by the CRIRES spectrograph \citep{kaeufl2004} at the VLT. These were obtained as part of the CRIRES survey of hot Jupiter atmospheres \citep{snellen2011_survey} first presented in \citet{schwarz2015} and more recently reanalysed in \citet{brogi2017, hawker2018} and \citet{brogi2019}. These observations cover the 2.29-2.35~$\mu$m range, where H$_2$O and CO show strong absorption (see Figure \ref{fig:cs}). The HRS data analysis on these high-resolution observations is discussed in section \ref{sec:hiresobs}.

\subsection{Likelihood Calculation}
\label{sec:likelihoodcombo}
Once the low- and high-resolution models have been computed, their log-likelihoods are calculated relative to the low- and high-resolution data, respectively. The overall likelihood for the atmospheric model is then found by taking the product of the low- and high-resolution likelihoods. The log-likelihood for the model comparison with LRS data is based on the chi-square metric, as has been used in many previous works \citep[e.g.][]{gandhi2018,brogi2019}:
\begin{equation*}
    \ln(\mathcal{L}_{\mathrm{low}}) =- \sum_{i}\frac{(y_{\mathrm{obs},i}-y_{\mathrm{model},i})^2}{2\sigma_i^2},
\end{equation*}
where the sum is over all data points, $y_{\mathrm{obs},i}$ are the observed fluxes with uncertainties $\sigma_i$, and $y_{\mathrm{model},i}$ are the corresponding model fluxes.

Since the high-resolution Doppler spectroscopy observations are analysed using cross-correlation, a different metric is used to assess the goodness of fit. As discussed by \citet{brogi2019}, such a metric should preserve the sign of the correlation coefficient such that emission lines are not fitted by absorption lines (and vice versa), and the metric should be sensitive to the scaling of the model. In this work, we use the metric developed by \citet{brogi2019} (discussed in section \ref{sec:HRCCloglikelihood}) to calculate the high-resolution log-likelihood, $\ln(\mathcal{L}_{\mathrm{high}})$. The total log-likelihood for a given atmospheric model is then
\begin{equation*}
    \ln(\mathcal{L}_{\mathrm{tot}}) = \ln(\mathcal{L}_{\mathrm{low}}) + \ln(\mathcal{L}_{\mathrm{high}}).
\end{equation*}

\section{high-resolution Observations}\label{sec:hiresobs}

Here we describe the data analysis of the HRS observations. The data were taken over 2 nights (18 and 25 July 2011) across the 2.29-2.35~$\mu$m K-band \citep{snellen2011_survey} with a spectral resolution R$\sim10^5$. There are 59 and 54 spectra from the first and second night respectively. Both sets of spectra are phase resolved across a range of $\phi\sim$ 0.5 - 0.6 capturing bright dayside emission of HD~209458b just after secondary eclipse. This dataset was first used to provide evidence for CO in the dayside atmosphere \citep{schwarz2015} and subsequent studies combining low-resolution observations detected CO \citep{brogi2017, brogi2019}. A more recent analysis of the high-resolution data set has also resulted in the detection of CO and H$_2$O \citep{hawker2018}, though we note that detecting H$_2$O is somewhat sensitive to the detrending procedure for removing telluric lines. We use the same data reduction procedure as that described in \citet{hawker2018} and \citet{cabot2019}.

\begin{figure}
    \centering
    \includegraphics[width=0.47\textwidth]{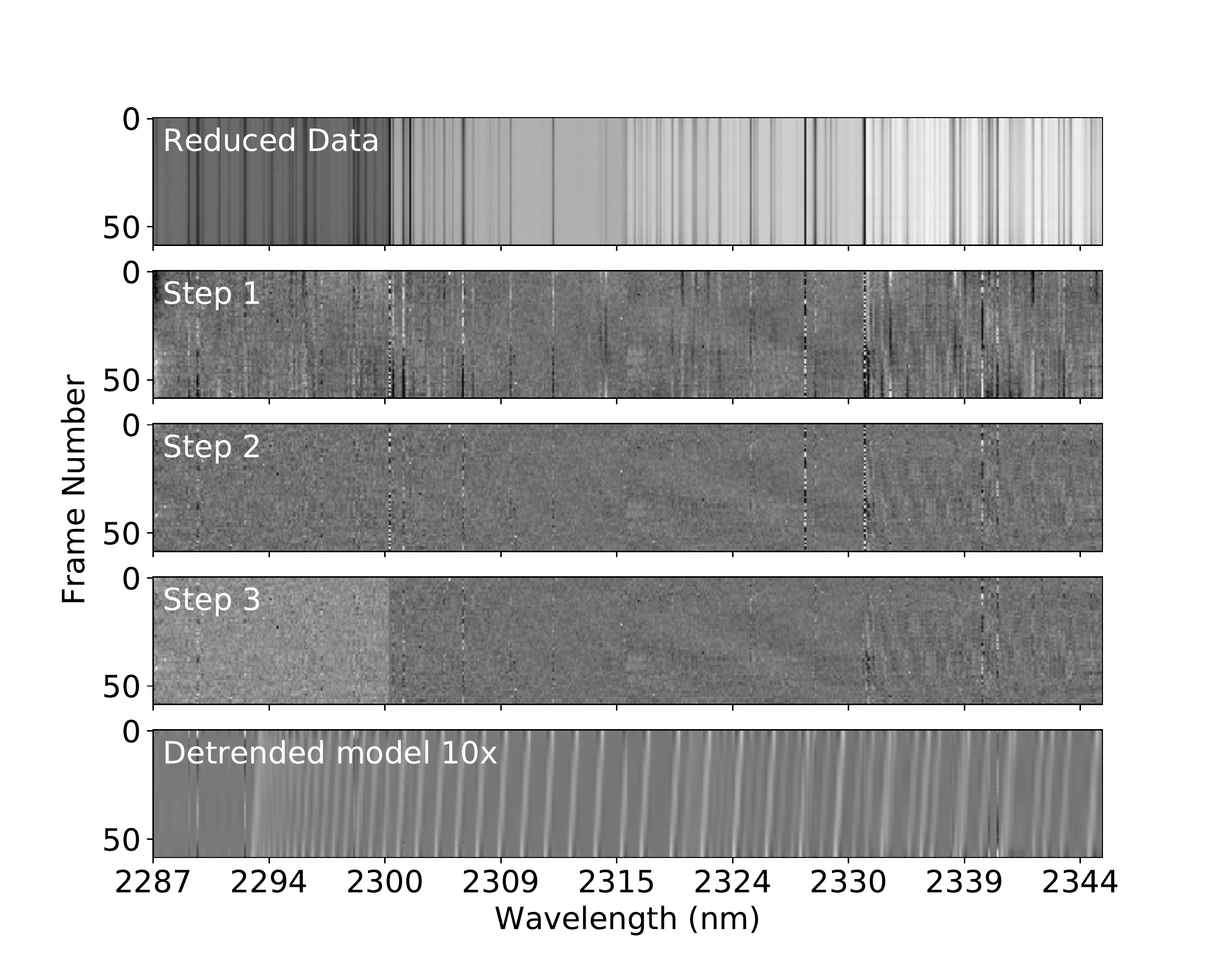}
    \caption{Illustration of the detrending steps (detailed in section \ref{sec:hiresobs}.2) as performed on the K-band CRIRES observations from 18 July 2011 with the data from all 4 detectors shown. The bottom panel shows a 10x strength model constructed prior to cross correlation (see section \ref{sec:HRCCloglikelihood}) treated with same detrending method. The noisy channel ($\sim 2332$nm) seen in steps 1 and 2 is from a detector edge and is dealt with by masking in step 3.}
    \label{fig:detrending}
\end{figure}

\subsection{Detrending}
Once reduced, the spectra from each detector form an M by N matrix where M is the number of frames from a given observation and N is the number of wavelength channels, typically 1024. The flux in a given pixel, F, is a function of wavelength and time given by the sum of the stellar (F$_s$) and planetary flux (F$_p$) multiplied by the telluric absorption (T$_E$).
\begin{equation}
    F(\lambda, t) = T_\mathrm{E}(\lambda, t) F_\mathrm{s}(\lambda, t) \Big(1 + \frac{F_\mathrm{p}(\lambda, t)}{F_\mathrm{s}(\lambda, t)}\Big)
    \label{eq:data_eq}
\end{equation}
The telluric absorption is expected to be constant with time and stellar flux quasi-constant with time as the stellar motion is on the order of ms$^{-1}$ with negligible Doppler shift compared to the planet motion ($\sim$kms$^{-1}$) which shifts the planet signal across many wavelength channels during a night of observations. Using the same procedure outlined in \citet{brogi2019}, we remove the stellar and telluric features through the steps illustrated in Figure \ref{fig:detrending} and described as follows:
\begin{itemize}
    \item[1.] We find the mean spectrum over time and fit it to each individual spectrum with a 2nd order polynomial and divide out the fit.
    \item[2.] We remove significant time-dependent residuals by fitting a 2nd order polynomial to each wavelength channel and divide out the fit. Such residuals are associated with telluric H$_2$O lines.
    \item[3.] We mask any noisy wavelength channels and update N; these masked noisy channels have a standard deviation $>$3 times the standard deviation of the matrix . The data corresponding to each spectrum are also mean subtracted, removing the continuum prior to cross-correlation. 
\end{itemize}

In essence the detrending process uses the data to find an approximation of the multiplicative term $T_\mathrm{E}(\lambda, t) F_\mathrm{s}(\lambda, t)$ in equation \ref{eq:data_eq} and divide it out to leave the planet to star flux ratio (plus one). In reality the multiplicative term is not perfectly removed hence the need to detect the planet via cross-correlation of the detrended data with models. We use this detrending method, however, for the following reasons. Firstly, we can validate our hybrid retrieval framework through comparison of our high-resolution results with \citet{brogi2019}. Secondly, it is computationally straightforward and has relatively few parameters (e.g. order of each polynomial fit, and n-sigma masking). Thus, during the retrieval it can be applied to each of the tested models to fairly account for the effects of the detrending process on the planet signal to avoid biases. The potential limitation of the method is whether it approximates the $T_\mathrm{E}(\lambda, t) F_\mathrm{s}(\lambda, t)$ term as well as other more computationally expensive algorithms such as SYSREM or PCA, potentially preventing it from being as effective. This is especially pertinent for analysis of more heavily contaminated spectral ranges such as the 3~$\mu$m L-band. Further discussions and comparisons on detrending methods can be found in \citet{hawker2018} and \citet{cabot2019}. 

\subsection{The HRS Log-likelihood Calculator}
\label{sec:HRCCloglikelihood}

As noted above, the detrending process will alter the planet signal. Thus, prior to cross-correlation, we treat each model in as similar way as possible to real data. A high-resolution model spectrum (generated by the architecture detailed in section \ref{sec:methods}) is fed to the HRS likelihood calculator along with the retrieval parameters d$V_{\mathrm{sys}}$ and d$K_\mathrm{p}$. The model spectrum is convolved to the CRIRES spectral resolution R$\approx 84,000$, then Doppler shifted using linear interpolation by velocities calculated according to equation \ref{eq:orbitalv} to produce a model equivalent of the $F_\mathrm{p}/F_\mathrm{s}$ term in equation \ref{eq:data_eq} as an M by N matrix.
\begin{equation}
    V_\mathrm{p}(t) = (K_\mathrm{p} + \mathrm{d}K_\mathrm{p})\sin(2\pi\phi(t)) + V_{\mathrm{sys}} + \mathrm{d}V_{\mathrm{sys}} + V_\mathrm{b}(t)
    \label{eq:orbitalv}
\end{equation}
The model $(1+F_\mathrm{p}/F_\mathrm{s})$ is then multiplied by the $T_\mathrm{E}F_\mathrm{s}$ matrix obtained from and divided out of the data during detrending. The detrending steps 1-3 are then applied to the model $T_\mathrm{E}F_\mathrm{s}(1+F_\mathrm{p}/F_\mathrm{s})$, similar to detrending the observations. The detrended model is then cross correlated with the detrended data. The likelihood is calculated via the likelihood mapping derived in \citet{brogi2019} which uses \citet{zucker2003} as a base and is summarised as follows. Each detrended observed spectrum $s(\lambda)$ is modelled by a detrended model spectrum $m(\lambda)$ Doppler shifted in wavelength by $\Delta\lambda$ and scaled by a factor $a$, added with some Gaussian noise with a standard deviation $\sigma$($\lambda$) according to equation \ref{eq:model}.
\begin{equation}
    s(\lambda) = a m(\lambda-\Delta \lambda) + \sigma(\lambda)  \mathcal{N}(0,1)
    \label{eq:model}
\end{equation}

The log-likelihood for such a model is given by equation \ref{eq:logL1}, where N is the number of wavelength channels.
\begin{equation}
    \log(\mathcal{L}_{\mathrm{high}}) = -N\log(\sigma(\lambda)) - \frac{1}{2\sigma^2}\sum_{\lambda}[s(\lambda)-am(\lambda-\Delta\lambda)]^2
    \label{eq:logL1}
\end{equation}
The maximum likelihood estimator $\hat{\sigma}$ is found by setting the partial derivative $\partial_{\sigma} \log(L)=0$. Setting $ \sigma = \hat{\sigma} $ in equation \ref{eq:logL1} and discarding the additive constant $N/2$, gives equation \ref{eq:logL2}.
\begin{equation}
    \begin{split}
    \log(\mathcal{L}_{\mathrm{high}}) = - \frac{N}{2}\log\Big[\frac{1}{N}\sum_{\lambda}[& s(\lambda)^2 - 2s(\lambda)am(\lambda-\Delta\lambda) \\ + & a^2m(\lambda-\Delta\lambda)^2]\Big]
    \end{split}
    \label{eq:logL2}
\end{equation}
A final detail is in setting the scale factor $a$. Mathematically, the maximum likelihood estimator can be used and found from $\partial_{a} \log(L)=0$. A more physically motivated approach would be to set $a=1$ to avoid unphysical scalings of the planet to star flux ratio and line strengths. In a HRS retrieval, imposing $a=1$ ensures the line depths of the tested models are set by the physical/chemical properties of the atmospheric model. In the hybrid retrieval including $a$ as a retrieval parameter is useful as a method of ensuring the P-T profile is correctly constrained by the low-resolution data. This is because the HRS has a tendency to favour stronger line strengths corresponding to higher temperatures but given the lack of continuum the high-resolution data should not be used to constrain the absolute temperature thus using $a$ makes an allowance for this effect. By performing the summation and setting a=1, the meaning of the terms inside the $\log$ of equation \ref{eq:logL2} becomes clear (see equation \ref{eq:logL3}). 

\begin{equation}
    \log(\mathcal{L}_{\mathrm{high}})= -\frac{N}{2}\log\Big(\langle s^2\rangle_{\lambda} -2 \langle sm\rangle_{\lambda} +\langle m^2\rangle_{\lambda}\Big)
    \label{eq:logL3}
\end{equation}
The first and third terms are the variances of a given detrended observed spectrum and detrended model spectrum respectively. The second term is twice the normalised dot product of the detrended observed spectrum with the detrended model spectrum and calculated when the two are cross correlated. Whilst the first term is a constant, the second and third terms are model dependent and also vary with $\Delta\lambda$ (a proxy for d$V_{\mathrm{sys}}$ and d$K_\mathrm{p}$). Using this method, every spectrum from a given night and detector yields a log-likelihood which are summed for all spectra to obtain a single log-likelihood from the high-resolution analysis. This log-likelihood can then be used to either perform a standalone HRS retrieval or used in a joint LRS+HRS retrieval by simply summing the log-likelihoods produced from the two analyses. As detailed in section \ref{sec:methods} Bayesian analysis can then be performed to retrieve the parameters of interest. For additional details about the likelihood mapping we refer the reader to \citet{brogi2019} and \citet{zucker2003}.

\section{Validation}
\label{sec:validation}

\begin{figure*}
\centering
\begin{overpic}[width=0.95\textwidth]{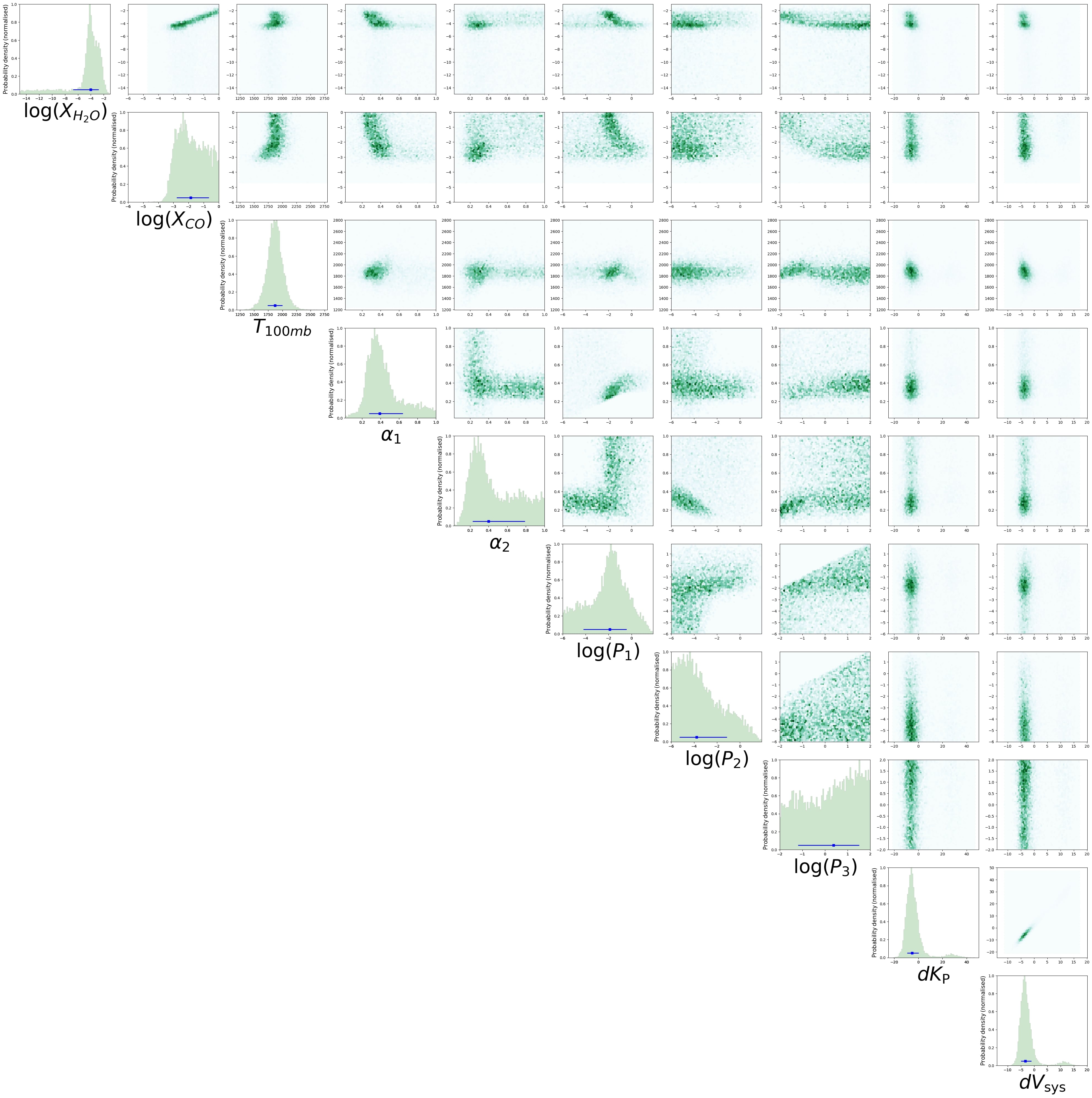}\large
\put (-5,31) {\def\arraystretch{1.3}
\begin{tabular}{l|{c}r}
\textbf{Parameter}              & \textbf{Value} & \textbf{Error} \\
\hline
log(X$_\mathrm{H_2O}$) & -4.0 & $\substack{+1.2 \\ -2.7}$ \\
log(X$_\mathrm{CO  }$) & -1.9 & $\substack{+1.2 \\ -0.9}$  \\
T$_\mathrm{100mb}$ /K     & 1869 & $\substack{+136 \\ -133}$ \\
$\alpha_1 \, /\mathrm{K}^{-\frac{1}{2}}$     & 0.39 & $\substack{+0.25 \\ -0.12}$ \\
$\alpha_2 \, /\mathrm{K}^{-\frac{1}{2}}$     & 0.40 & $\substack{+0.39 \\ -0.17}$ \\
log(P$_1$/bar)     & -1.9 & $\substack{+1.5 \\ -2.3}$ \\
log(P$_2$/bar)     & -3.8 & $\substack{+2.6 \\ -1.5}$ \\
log(P$_3$/bar)     & 0.4 &$\substack{+1.1 \\ -1.6}$\\
d$K_\mathrm{p}$   \, /kms$^{-1}$    & -5.2        & $\substack{+5.5 \\ -4.1}$\\ 
d$V_{\mathrm{sys}}$ \, /kms$^{-1}$   & -3.3        & $\substack{+2.3 \\ -1.6}$\\
\end{tabular}
}
\end{overpic}
    \caption{Posterior distribution of HD~209458b from the retrieval of the high-resolution 2.29-2.35~$\mu$m CRIRES observations \citep{snellen2011_survey}. We retrieved two volatile chemical species, H$_2$O and CO, and parametrised the atmospheric temperature profile with six parameters, as discussed in \citet{gandhi2018}. We also include two additional parameters, d$K_\mathrm{p}$ and d$V_\mathrm{sys}$.}
    \label{fig:posterior_validation}
\end{figure*}

\begin{figure*}
\centering
\begin{overpic}[width=0.95\textwidth]{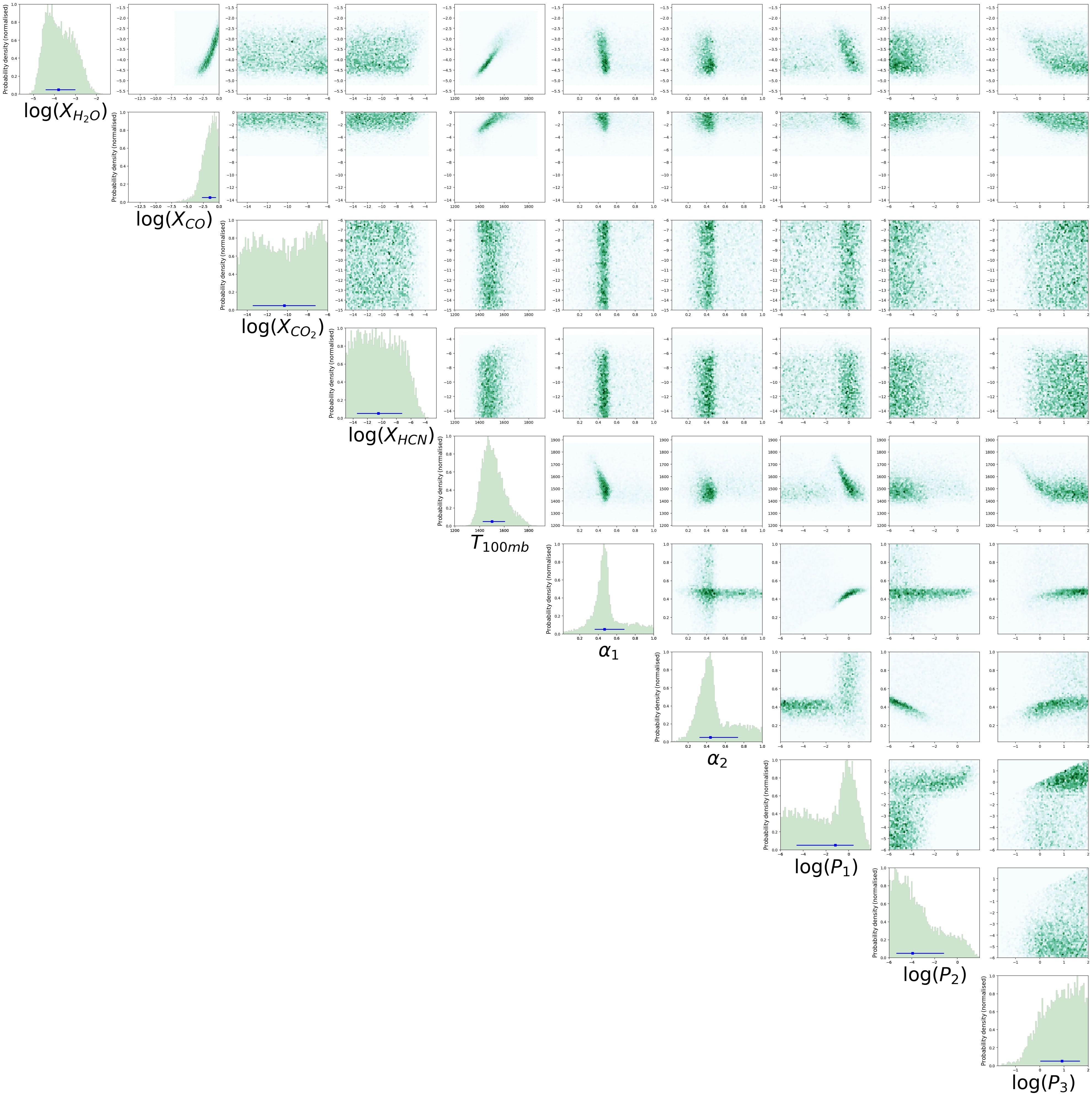}\large
\put (-5,31) {\def\arraystretch{1.3}
\begin{tabular}{l|{c}r}
\textbf{Parameter}              & \textbf{Value} & \textbf{Error} \\
\hline
log(X$_\mathrm{H_2O}$) & -3.81 & $\substack{+0.8 \\ -0.62}$ \\
log(X$_\mathrm{CO  }$) & -1.4 & $\substack{+1.0 \\ -1.3}$  \\
log(X$_\mathrm{CO_2  }$) & <-6.2 & -  \\
log(X$_\mathrm{HCN  }$) & <-5.5 & -  \\
T$_\mathrm{100mb}$ /K     & 1501 & $\substack{+106 \\ -75}$ \\
$\alpha_1 \, /\mathrm{K}^{-\frac{1}{2}}$     & 0.47 & $\substack{+0.22 \\ -0.11}$ \\
$\alpha_2 \, /\mathrm{K}^{-\frac{1}{2}}$     & 0.44 & $\substack{+0.30 \\ -0.12}$ \\
log(P$_1$/bar)     & -1.2 & $\substack{+1.6 \\ -3.4}$ \\
log(P$_2$/bar)     & -3.9 & $\substack{+2.8 \\ -1.4}$ \\
log(P$_3$/bar)     & 0.92 &$\substack{+0.74 \\ -0.90}$\\
\end{tabular}
}
\end{overpic}
    \caption{Posterior distribution of HD~209458b from the retrieval of the low-resolution HST WFC3 and Spitzer observations \citep{line2016, diamondlowe2014}. We retrieved four volatile chemical species, H$_2$O, CO, CO$_2$ and HCN, and parametrised the atmospheric temperature profile with six parameters, as discussed in \citet{gandhi2018}.}
    \label{fig:posterior_lr}
\end{figure*}

We now validate our hybrid retrieval framework HyDRA-H by comparing to previous work \citep{brogi2019} which uses HRS-only observations. We retrieve the atmospheric properties of HD~209458b using high-resolution spectra of the system observed over two nights at phases of $\phi\sim$ 0.5 - 0.6, i.e. shortly after secondary eclipse. As discussed in section \ref{sec:hiresobs}, we use archival observations obtained as part of the CRIRES survey of hot Jupiters \citep{snellen2011_survey}. These spectra span the 2.29-2.35~$\mu$m wavelength range and have been obtained at a resolution of R$\sim 10^5$. Following \citet{brogi2019} we retrieve the H$_2$O and CO volume mixing ratios, d$K_\mathrm{p}$, d$V_\mathrm{sys}$ and the temperature profile. We parametrise the P-T profile over six free parameters, T$_\mathrm{100mb}$, $\alpha_1$, $\alpha_2$, P$_1$, P$_2$ and P$_3$ using the prescription of \citet{madhu2009}. We run our retrievals using Nested Sampling with 1000 live points and $\sim$22,000 wavelength points at 0.01cm$^{-1}$ spacing between 2.25-2.37~$\mu$m. This corresponds to a model spectral resolution of R$\sim 4\times 10^5$. We conduct the retrievals using the high-resolution observations only as was done in \citet{brogi2019}.

Figure \ref{fig:posterior_validation} shows the posterior distributions for the retrieved parameters. The P-T profile is shown in Figure \ref{fig:PT} and agrees well with the constrained P-T profile from the high-resolution retrieval by \citet{brogi2019}. The planet's photospheric temperature is also within 1$\sigma$ of the equilibrium temperature and shows no signs of a thermal inversion.

The deviations d$K_\mathrm{p}$ and d$V_\mathrm{sys}$ are within $\sim$1$\sigma$ of the expected value of 0, indicating that the planet's signal has been detected at the expected location in the $K_\mathrm{p}-V_\mathrm{sys}$ plane without any spurious peaks (e.g from stellar lines) interfering with the result significantly. We do observe a very weak peak at positive d$K_\mathrm{p}$ and d$V_\mathrm{sys}$ values which we attribute to noise in the $K_\mathrm{p}-V_\mathrm{sys}$ plane. Figure \ref{fig:posterior_validation} shows that the two parameters are correlated with each other in a similar way to that seen in previous high-resolution studies which perform cross correlation on the spectral models \citep{schwarz2015, hawker2018}.

We also find good agreement in the retrieved chemical abundances of H$_2$O and CO to both previous high-resolution and low-resolution retrievals \citep{line2016, brogi2019}. We constrain the H$_2$O to be $\log(\mathrm{H_2O)} = -4.0^{+1.2}_{-2.7}$, consistent to within $\sim$1$\sigma$ of \citet{brogi2019} as shown in the HRS-only posterior in Figure \ref{fig:histogram}. This abundance estimate is also consistent with expectations from chemical equilibrium at solar abundance \citep{madhu2012, moses2013} but also allows for significantly sub-solar H$_2$O. The retrieval does indicate a long tail in the distribution at low abundance for the H$_2$O. Hence our detection significance for H$_2$O in this HRS-only retrieval is lower than the other retrievals we conduct (see Table~\ref{tab:detection_significances}). We are not able to strongly constrain the H$_2$O given that its features are weak in this spectral range. On the other hand, CO has a strong cross section in this part of the spectrum (see Figure \ref{fig:cs}) and is therefore better constrained at $\log(\mathrm{CO)} = -1.9^{+1.2}_{-0.9}$. The CO thus has a higher detection significance than the H$_2$O in the high-resolution only retrieval at 4.6$\sigma$ confidence because of its stronger cross section. This result is also consistent with \citet{brogi2019}. In section \ref{sec:results} we will compare and contrast these to the low-resolution and hybrid constraints.

Previous work has shown that the choice of line list for H$_2$O can influence detections and the abundance estimates \citep{brogi2019}. Thus we have used line lists for H$_2$O and CO from the HITEMP database \citep{rothman2010} as used by \citet{brogi2019}. Our temperature profile on the other hand has been parametrised differently to \citet{brogi2019}, but ultimately the P-T profiles are in agreement to within 1$\sigma$, particularly for the photosphere at P$\sim$0.1~bar. We are able to constrain the temperature profile similarly despite the differences in parametrisation as both allow a full exploration of the possible parameter space, encompassing thermal inversions and non-inverted profiles.

Whilst we generally observe good agreement with previous work by \citet{brogi2019}, we do note some differences in the posteriors for the HRS-only retrieval. Whilst consistent to within 1$\sigma$, our abundance peak for H$_2$O is at a slightly higher abundance and our CO abundance only shows a strong lower limit. Previous high resolution analysis using a different method also indicated sub-solar H$_2$O \citep{brogi2017} whereas we see an H$_2$O peak at a sub-solar value but which allows for solar values as well. This difference may be attributed to some differences in our models and/or data analysis. Our P-T profile is parameterised from the work in \citet{madhu2009}, whereas \citet{brogi2019} use the \citet{guillot2010} prescription. We see a slightly shallower temperature gradient in the atmosphere which allows the abundance of H$_2$O and CO to extend to higher values to explain the line depths seen in the observations. However, this difference is less of a concern as the LRS observations are much more sensitive to the temperature profile than HRS and thus give better constraints for the hybrid and LRS-only retrieval as shown in the posteriors in Figures \ref{fig:posterior_lr} and \ref{fig:posterior} and discussed in section \ref{sec:results}. We also see some differences in the retrieved values of d$K_\mathrm{p}$ and d$V_\mathrm{sys}$. These likely arise from our analysis of the CRIRES data. Our wavelength calibration is performed independently and our orbital solution is potentially different to that used in \citet{brogi2019} which would affect d$K_\mathrm{p}$ and d$V_\mathrm{sys}$. This could also slightly alter the retrieved abundances. Overall, the differences in the retrieved parameters are small and show good agreement with \citet{brogi2019} and expected values from known orbital parameters.

With this we are now ready to perform the hybrid retrieval of HD~209458b using both the low-resolution and the high-resolution observations simultaneously. We include three additional parameters for our hybrid retrieval, namely the HCN and CO$_2$ volume mixing ratios, and the high-resolution model scale factor $\log(a)$ as included for the simulated retrievals in \citet{brogi2019}. We include H$_2$O, CO and HCN as there has been evidence for their presence in hot Jupiter atmospheres from previous analyses of high-resolution infrared spectra \citep[e.g.][]{birkby2013, schwarz2015, brogi2017, hawker2018, cabot2019}. The CO$_2$ has also been shown to have a degeneracy with CO in the 4.5~$\mu$m Spitzer band \citep{madhu2010, line2016, gandhi2018} and thus we include it to compare any degeneracies which may arise. We include the scale factor $\log(a)$ as a way for the hybrid retrieval to weight the likelihoods independently from each set of observations following \citet{brogi2019}.

\section{Results}
\label{sec:results}

In this section we discuss the results from our simultaneous hybrid retrieval of the low-resolution and high-resolution spectra of HD~209458b in thermal emission. The HRS data in the 2.29-2.35~$\mu$m range are described in section \ref{sec:validation}. The LRS data in the HST WFC3 band (1.1-1.7~$\mu$m) are obtained from \citet{line2016} and the Spitzer observations in the $\sim$3-10~$\mu$m range are obtained from \citet{diamondlowe2014}. The HyDRA-H retrieval framework developed to perform the simultaneous hybrid retrieval is discussed in section \ref{sec:methods}. We will compare and contrast the results from the hybrid retrieval with the HRS-only retrieval discussed in section \ref{sec:validation} as well a LRS-only retrieval.

\begin{figure}
\centering
	\includegraphics[width=\columnwidth,trim={0.8cm 0 2.3cm 0},clip]{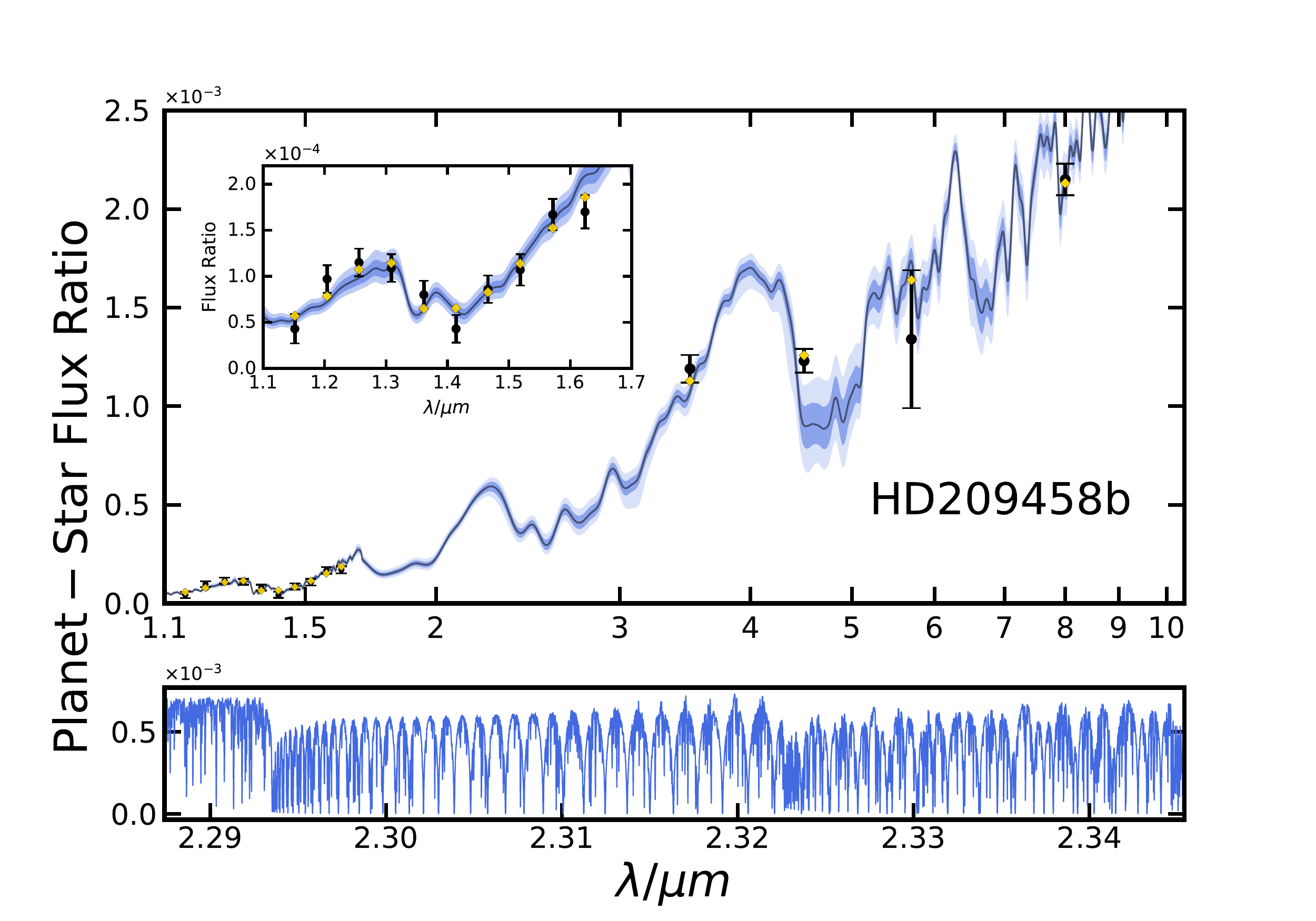}
    \caption{Thermal emission spectrum of of HD~209458b. The black data points in the top panel show the low-resolution HST and Spitzer observations of the planet-star flux ratio \citep{line2016} with their associated error bar. The inset shows the 1.1-1.7~$\mu$m HST Wide Field Camera 3 region. The dark and light shaded regions show the 1$\sigma$ and 2$\sigma$ uncertainties respectively from the retrieval models and the gold markers indicate the binned data points from the median model. The bottom panel shows the best fit spectrum from the hybrid retrieval in the observed 2.29-2.35~$\mu$m spectral range of the high-resolution observations.}     
\label{fig:flux}
\end{figure}

\begin{figure*}
\centering
\begin{overpic}[width=0.95\textwidth]{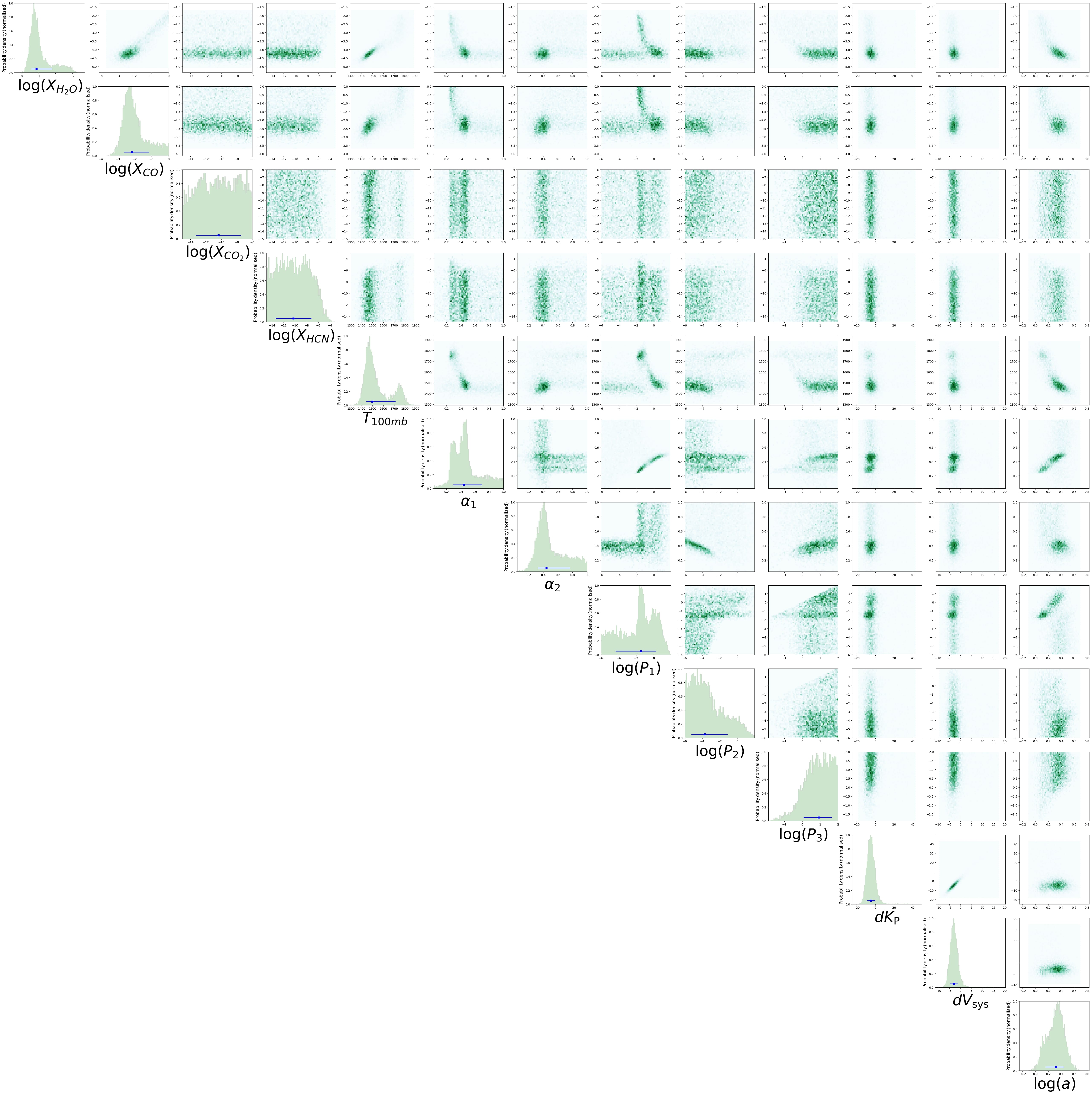}\large
\put (-5,31) {\def\arraystretch{1.2}
\begin{tabular}{l|{c}r}
\textbf{Parameter}              & \textbf{Value} & \textbf{Error} \\
\hline
log(X$_\mathrm{H_2O}$) & -4.11 & $\substack{+0.91 \\ -0.30}$ \\
log(X$_\mathrm{CO  }$) & -2.19 & $\substack{+0.99 \\ -0.47}$  \\
log(X$_\mathrm{CO_2}$) & < -6.3 & -  \\
log(X$_\mathrm{HCN }$) & < -5.4 & -  \\
T$_\mathrm{100mb}$ /K     & 1498 & $\substack{+216 \\ -57}$ \\
$\alpha_1 \, /\mathrm{K}^{-\frac{1}{2}}$     & 0.44 & $\substack{+0.25 \\ -0.15}$ \\
$\alpha_2 \, /\mathrm{K}^{-\frac{1}{2}}$     & 0.44 & $\substack{+0.32 \\ -0.11}$ \\
log(P$_1$/bar)     & -1.5 & $\substack{+1.7 \\ -2.9}$ \\
log(P$_2$/bar)     & -3.7 & $\substack{+2.6 \\ -1.5}$ \\
log(P$_3$/bar)     & 0.92 &$\substack{+0.74 \\ -0.84}$\\
d$K_\mathrm{p}$   \, /kms$^{-1}$    & -4.9        & $\substack{+4.4 \\ -4.1}$\\ 
d$V_{\mathrm{sys}}$ \, /kms$^{-1}$   & -3.0        & $\substack{+1.8 \\ -1.6}$\\
$\log(a)$          & 0.32 & $\substack{+0.12 \\ -0.16}$\\
\end{tabular}
}
\end{overpic}
    \caption{Posterior distribution of the hybrid retrieval of HD~209458b's dayside. The low-resolution dataset was obtained from \citet{line2016} and \citet{diamondlowe2014} and considers HST WFC3 and four Spitzer photometric channels between 3.6-10~$\mu$m and the high-resolution data was obtained from the CRIRES spectral survey \citep{snellen2011_survey}. We retrieved four volatile chemical species, H$_2$O, CO, CO$_2$ and HCN, and parametrised the atmospheric temperature profile with six parameters, as discussed in \citet{gandhi2018}. We also include two additional parameters for the high-resolution observations, d$K_\mathrm{p}$, d$V_\mathrm{sys}$ and $\log(a)$ as discussed in section \ref{sec:HRCCloglikelihood}.}
    \label{fig:posterior}
\end{figure*}

Figure \ref{fig:flux} shows the planet/star flux ratio for the best-fit high-resolution model as well as the spectral fit to the low-resolution observations from the hybrid retrieval. We see a good fit to the HST WFC3 and Spitzer observations, with a clear absorption feature in the WFC3 observations at $\sim$1.4~$\mu$m due to the presence of H$_2$O in the atmosphere. The Spitzer observations for the 4.5~$\mu$m band also show a feature due to CO. The high-resolution spectral model at R$\sim$400,000 in the 2.29-2.35~$\mu$m range also shows clear absorption features due to the presence of both H$_2$O and CO due to both species having strong cross sections in this band (see Figure \ref{fig:cs}). The constraints on these species are given in the histograms in Figure \ref{fig:histogram} and the full posterior distributions for all of the parameters are shown in Figure \ref{fig:posterior}.

Our results for the abundances are in good agreement with previous low-resolution and high-resolution retrievals on HD~209458b \citep{line2016, brogi2017, brogi2019}. The posterior distribution for the low-resolution retrieval that we performed with HyDRA \citep{gandhi2018} is shown in Figure~\ref{fig:posterior_lr}. The H$_2$O and CO abundances have been well constrained (Figure \ref{fig:histogram}) and are consistent with previous work \citep{line2016, brogi2019}. In addition, the deviations from the known planetary and systemic velocities, d$K_\mathrm{p}$ and d$V_\mathrm{sys}$, and the scale factor $\log(a)$ are also consistent with 0 within 2$\sigma$, as expected . This further validates that the high-resolution observations are able to constrain the planetary signal at the known location and strength without spurious detections. We will now further discuss each of these parameters below and compare and contrast the results with those obtained from the separate retrievals with the individual HRS vs. LRS datasets.

\subsection{H$_2$O Abundance}

\begin{figure*}
	\includegraphics[width=\textwidth,trim=1cm 0 1cm 0,clip]{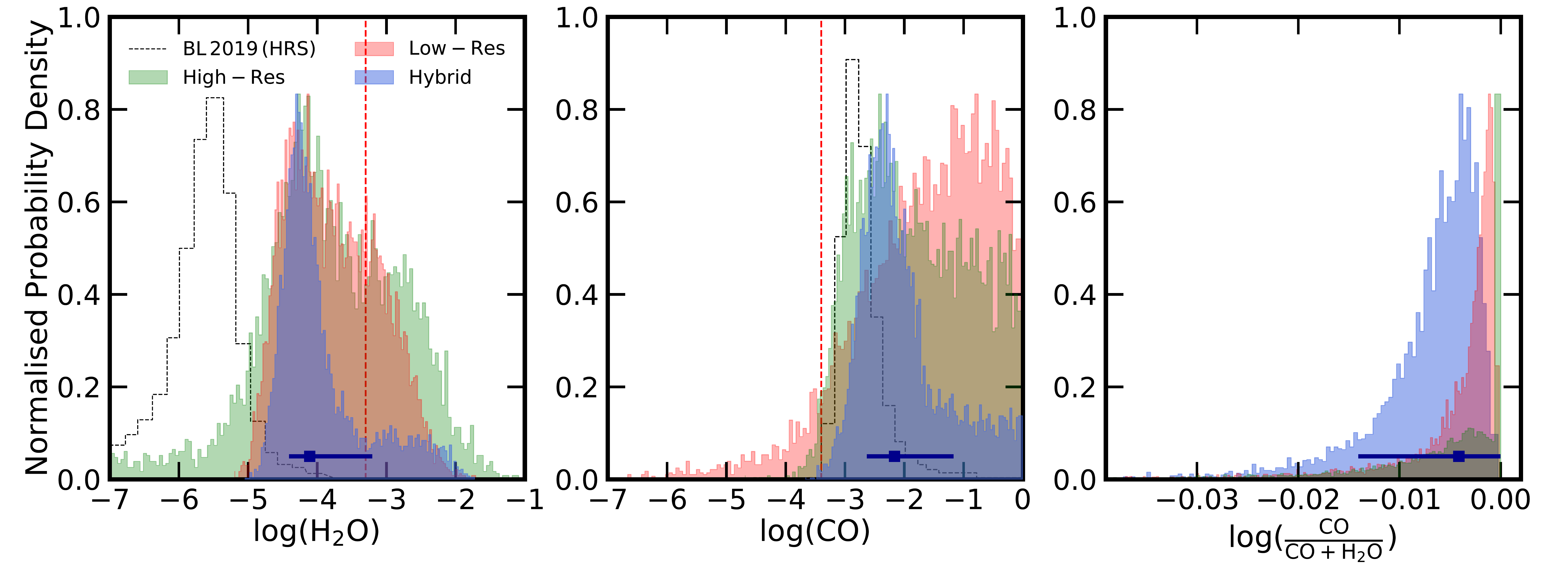}
    \caption{Probability distributions for the retrieved volume mixing ratios of H$_2$O and CO for the three retrievals run for HD~209458b. The left and centre panels show the H$_2$O and CO abundances respectively and the right panel shows the retrieved CO/(H$_2$O+CO) ratio. The dashed black lines show the posterior distributions for H$_2$O and CO from the retrieval of the high resolution data in \citet{brogi2019}. The error bar in each panel indicates the median and 1$\sigma$ uncertainty from the hybrid retrieval discussed in section \ref{sec:results}. The red dashed lines are the chemical equilibrium solar values at the planet equilibrium temperature and 0.1~bar pressure \citep{madhu2012}.}
    \label{fig:histogram}
\end{figure*}

Figure \ref{fig:histogram} shows the H$_2$O abundance for the hybrid retrieval as well as the LRS-only and HRS-only retrievals. The hybrid retrieval is able to better constrain the abundances as it utilises both the lower resolution HST as well as the high-resolution CRIRES observations. Our detection significance of 7.3$\sigma$ is also improved with the hybrid retrieval as shown in Table~\ref{tab:detection_significances}. The hybrid H$_2$O constraint is $\log(\mathrm{H_2O)} = -4.11^{+0.91}_{-0.30}$, which lies within each of the HRS- and LRS-only constraints. This abundance is also consistent with that seen from previous retrievals \citep{line2016, brogi2017, brogi2019}. While being consistent with solar to within 1$\sigma$, the H$_2$O abundance peaks at a sub-solar value of $\log(\mathrm{H_2O})\sim-4.2$.

The LRS-only retrieval is able to constrain H$_2$O due to the presence of an absorption feature in the HST WFC3 data. This is caused by the strong broad molecular cross section of H$_2$O at $\sim$1.4~$\mu$m as shown in Figure \ref{fig:cs}. Hence our detection significance of H$_2$O is 7.0$\sigma$ (see table~\ref{tab:detection_significances}). We constrain an H$_2$O abundance that is consistent with solar, but with a larger uncertainty than the hybrid retrieval. This result is consistent with previous low-resolution retrievals of the same dataset \citep{line2016} as well as previous retrievals of the planet with transmission spectra which indicate sub-solar abundances \citep{madhu2014, barstow2017, pinhas2019}.

The HRS data on the other hand are able to constrain and refine the H$_2$O estimate due to the numerous transition lines that H$_2$O possesses in the 2.29-2.35~$\mu$m spectral range. In this spectral range there are $\sim$1.6$\times$10$^6$ H$_2$O transition lines. The cross correlation is able to account for $\sim$10$^3$ of these transition lines which most strongly influence the spectrum (see Figure \ref{fig:flux}). However, given the weaker cross section of H$_2$O in this spectral range (see Figure \ref{fig:cs}) the abundance is not well constrained in this case compared with the other two retrievals. The detection significance of H$_2$O with the HRS-only retrieval is thus only 2.0$\sigma$. The uncertainty in the H$_2$O abundance is therefore also greatest when considering the HRS data alone. This is an important consideration which has also been discussed in previous work \citep{brogi2019}.

\begin{table}[ht]
    \centering
    \begin{tabular}{c|c|c}
        \textbf{Species} & \textbf{Retrieval} & \textbf{Detection Significance} \\
        \hline
        $\mathrm{H_2O}$ & Low-Resolution & 7.0$\sigma$\\
         & High-Resolution & 2.0$\sigma$ \\
         & \textbf{Hybrid} & \textbf{7.3}$\boldsymbol{\sigma}$ \\
        \hline
        $\mathrm{CO}$ & Low-Resolution & 2.6$\sigma$ \\
         & High-Resolution & 4.6$\sigma$ \\
         & \textbf{Hybrid} & \textbf{5.3}$\boldsymbol{\sigma}$ \\
    \end{tabular}
    \caption{Detection significances for H$_2$O and CO for each of the retrievals conducted.}
    \label{tab:detection_significances}
\end{table}

\subsection{CO Abundance}

Figure \ref{fig:histogram} shows the CO abundance constraints for the hybrid retrieval. The CO abundance is most strongly constrained from the Spitzer and HRS observations as CO does not have a strong cross section in the HST WFC3 range. Our detection significance for CO for the hybrid retrieval is 5.3$\sigma$ (see table~\ref{tab:detection_significances}). The hybrid constraints suggest a super-solar CO abundance of $\log(\mathrm{CO}) = {-2.16}^{+0.99}_{-0.47}$ compared to the solar value of $\log(\mathrm{CO}) \sim -3.4$ \citep{moses2013}. This, combined with the H$_2$O abundance, means that the C/O ratio for HD~209458b is obtained to be $\sim$1, consistent with that seen from previous high-resolution observations \citep{brogi2017, brogi2019}.

Our low-resolution retrieval constrains the CO due to the 4.5~$\mu$m Spitzer band where it has a strong cross section (see Figure \ref{fig:cs}). This data is only able to constrain a lower limit of $\log(\mathrm{CO}) \gtrsim -5$. In addition, the detection significance is weaker at 2.6$\sigma$, as shown in Table \ref{tab:detection_significances}. Whereas the H$_2$O had numerous HST WFC3 points to constrain its abundance, the CO is only constrained by the handful of Spitzer data points. Hence the CO is detected at lower significance for the low-resolution retrieval and has a wider abundance uncertainty than the H$_2$O.

The HRS observations on the other hand are able to constrain the CO by resolving the CO spectral lines in the 2.29-2.35~$\mu$m range. CO possesses strong opacity with $\sim$2800 transition lines in this range, of which $\sim$10$^2$ are strong enough to be cross correlated (see Figure \ref{fig:flux}). Despite being less numerous than the H$_2$O lines, the stronger cross section of CO in this spectral range mean that the high-resolution retrieval is able to detect and constrain CO despite the fewer transitions. Our results show that combining the high-resolution and the low-resolution observations provide a substantial improvement in constraining the CO abundance compared to either dataset alone.

\subsection{Other Species}

We retrieve CO$_2$ and HCN but we observe no constraint on either species using the current datasets as shown in the summary table in Figure \ref{fig:posterior}. The HCN does show an upper limit of $\log(\mathrm{HCN}) \lesssim -5.4$ at 2$\sigma$ confidence, but without any definitive detection it is difficult to further quantify this. HCN has previously been reported with HRS in the 3.18-3.27~$\mu$m CRIRES range \citep{hawker2018} where it has a strong cross section, but with only 2.29-2.35~$\mu$m observations in this work we are unable to see any significant constraints due to its weaker opacity compared with CO and H$_2$O (see Figure \ref{fig:cs}).

\subsection{C/O Ratio}

The right panel of Figure \ref{fig:histogram} shows the CO/(CO+H$_2$O) ratio. In the absence of any other species this may be used as a proxy for the C/O ratio, which we constrain to be $\log({\rm C/O}) = -0.0041^{+0.0041}_{-0.010}$. The corresponding C/O ratio is $0.991^{+0.009}_{-0.022}$, which is consistent with previous estimates using HRS spectra \citep{brogi2017, brogi2019}. This ratio is greater than the solar value of $\sim$0.54 \citep{asplund2009}, and is caused by the retrieved CO abundance being significantly higher than the H$_2$O, as shown in Figure \ref{fig:histogram}. A C/O$\sim$1 is also consistent with the recent inference of HCN using HRS data of HD~209458b in the 3.18-3.27~$\mu$m band \citep{hawker2018}, as well as with the low H$_2$O abundance reported in transmission \citep[e.g.,][]{madhu2014}. In chemical equilibrium HCN is only present at significant abundance at C/O ratios near unity \citep{madhu2012}. We should note however that the C/O ratio of the atmosphere may be influenced by other factors such as rain-out of oxygen bearing species and the presence of other C and O bearing gaseous chemical species.

\subsection{Temperature Profile}

\begin{figure}
\centering
	\includegraphics[width=\columnwidth,trim={0cm 0 1.4cm 0},clip]{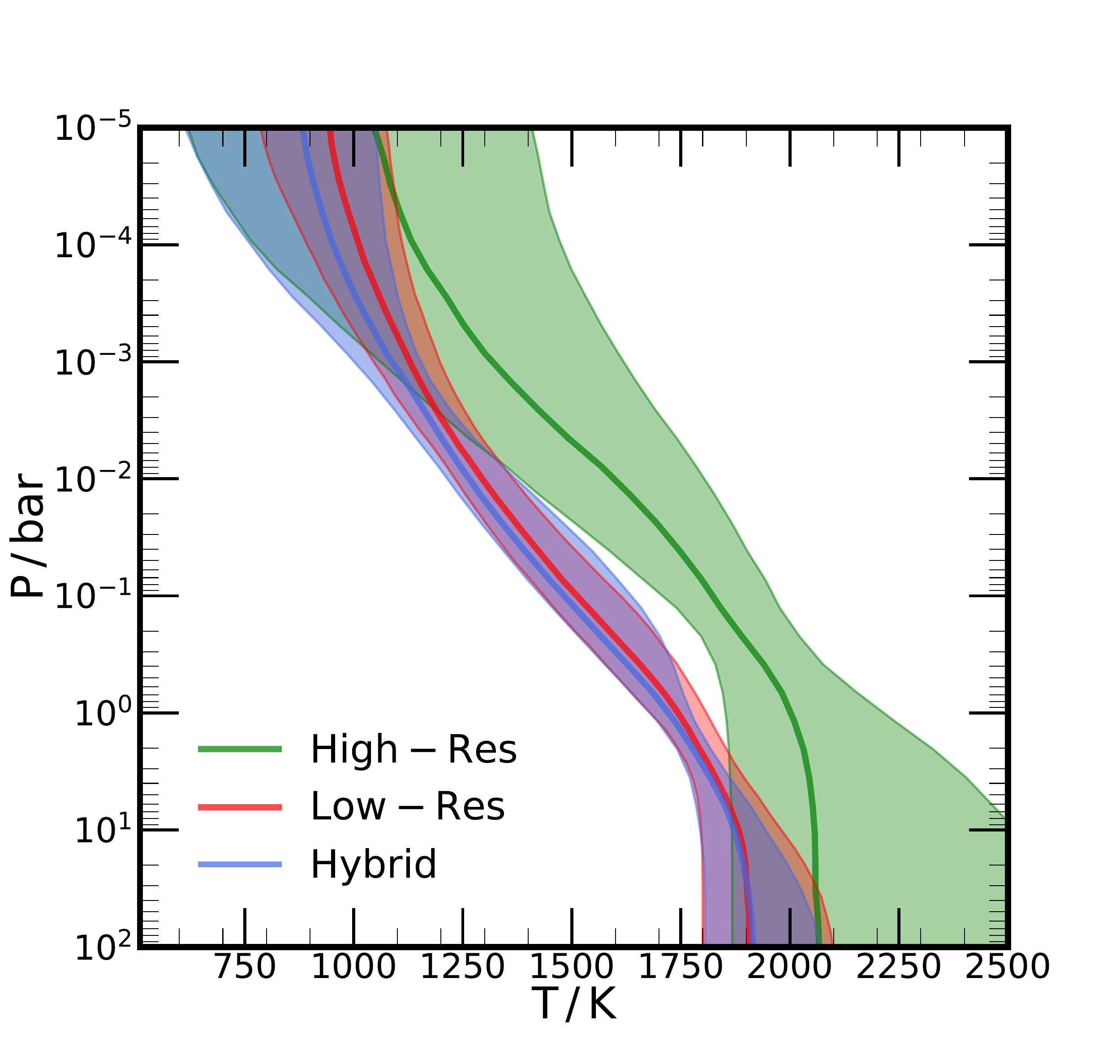}
    \caption{Retrieved atmospheric temperature profile from the three retrievals run for HD~209458b. The thick line and the shaded regions indicate the median P-T profile and 1$\sigma$ uncertainty respectively for each retrieval.}
\label{fig:PT}
\end{figure}

The P-T profile and the uncertainty are shown for all three retrievals in Figure \ref{fig:PT}. The majority of the temperature profile constraint occurs from the absorption features in the LRS data with some constraints from the absorption lines in the HRS data. These clearly indicate that the planet does not possess a thermal inversion (or stratosphere) within the photosphere, consistent with prevous studies \citep{line2016, brogi2019}. This is also expected given that species such as TiO or VO would have condensed out given that the planet's dayside temperature is below $\sim$1800~K \citep{spiegel2009}. The retrieved 100mb temperature is $1498^{+216}_{-57}$~K, consistent with the equilibrium temperature of $\sim$1450~K assuming full redistribution. 

From Figure \ref{fig:PT} we see that the tightest constraint on the temperature occurs in the photosphere at P$\sim$1~bar. At pressures greater than this the optical depth $\tau >> 1$ and thus does not significantly contribute to the overall spectrum. The onset of the isotherm at pressures P$\sim$1-10~bar is consistent with that seen from self-consistent equilibrium models of hot Jupiter atmospheres \citep{fortney2008, burrows2008, gandhi2017}. The high-resolution spectra probe the line centres which originate higher up in the atmosphere at lower pressures. However, given that the high-resolution observations are less sensitive to the temperature and more so to the gradient, the temperature constraint does not improve significantly over the low-resolution retrieval. Going to even lower pressures, the temperature is now less well constrained as the optical depth is too small to contribute significantly to the emergent spectrum even in the line cores. Therefore the uncertainty for the top of the atmosphere is much greater.

\subsection{High-Resolution Parameters}

In addition to the chemical abundances and temperature profile, we also retrieve the deviation from the known value of the planetary and systemic velocity and the scale factor. Both d$K_\mathrm{p}$ and d$V_\mathrm{sys}$ are consistent with 0 to within 1.5$\sigma$ as shown in the table in Figure \ref{fig:posterior}. The scale factor is also retrieved to be $\log(a) = 0.32^{+0.12}_{-0.16}$. These three additional parameters confirm that the high-resolution retrieval is able to retrieve the planetary signal without biasing the results or detecting spurious peaks in the $K_\mathrm{p}$-$V_\mathrm{sys}$ plane.

\section{Summary and Discussion}
\label{sec:conclusion}

In this work, we develop a new hybrid atmospheric retrieval code and perform the first simultaneous hybrid retrieval of low- and high-resolution data for the exoplanet HD 209458b. Our code, HyDRA-H, builds upon the hybrid retrieval framework of \citet{brogi2019} and is capable of retrieving atmospheric properties of exoplanets using low- and high-resolution emission spectra simultaneously. The architecture extends our recently developed HyDRA retrieval framework \citep{gandhi2018} by incorporating methods for retrieval of high-resolution spectra from \citet{brogi2019}. We also perform separate retrievals on the low-resolution and high-resolution data. These separate retrievals demonstrate the contribution of each approach to the constraints obtained in the simultaneous hybrid retrieval.

We have validated our HyDRA-H retrieval framework by comparing to previous high-resolution retrievals by \citet{brogi2019}. We retrieve the 2.29-2.35~$\mu$m VLT CRIRES observations of HD~209458b \citep{snellen2011_survey} and constrain the H$_2$O and CO volume mixing ratios and parametrised temperature profile. We additionally constrain the deviations from the known value of the planetary and systemic velocities, d$K_\mathrm{p}$ and d$V_\mathrm{sys}$. This is to confirm that the planetary signal is located in the expected position in the $K_\mathrm{p}-V_\mathrm{sys}$ plane and is not significantly affected by spurious signals such as those from the host star or telluric contamination. We find that the H$_2$O and CO abundances given in Figure \ref{fig:histogram} are in agreement with previous high-resolution retrievals \citep{brogi2017, brogi2019}. The temperature profile is also consistent with previous work and the photosphere temperature is consistent with the equilibrium temperature of HD~209458b.

The retrieval on the low-resolution data is able to constrain the molecular abundances and temperature profiles from the spectral features seen in the HST WFC3 and Spitzer observations. We constrain H$_2$O from its strong absorption feature in the WFC3 band and find it to be consistent with previous low-resolution retrievals of this dataset \citep{line2016}. Our CO abundance is constrained primarily by the 4.5~$\mu$m Spitzer point. Both of these chemical species are consistent with the high-resolution retrieval. The P-T profile for the low-resolution retrieval does however show some differences. The retrieval indicates a photosphere temperature that is $\sim$1.5$\sigma$ away from the high-resolution observations. However, this is not unexpected given that the high-resolution observations remove the continuum and thus an absolute temperature is more difficult to constrain.

Our hybrid retrieval is able to use the strengths of both high-resolution and low-resolution observations to constrain the abundances and the temperature profile. Hence we see a higher detection significance for each species, as shown in Table~\ref{tab:detection_significances}, than for the retrievals on the individual datasets separately. We therefore also see much tighter constraints on the H$_2$O and CO mixing ratios in Figure \ref{fig:histogram}. The hybrid retrieval is able to use the information from both available observations and combine their likelihoods to obtain more stringent constraints. We retrieve the H$_2$O abundance to be $\log(\mathrm{H_2O)} = -4.11^{+0.91}_{-0.30}$ and the CO to be $\log(\mathrm{CO}) = {-2.16}^{+0.99}_{-0.47}$. The H$_2$O is sub-solar but consistent to within 1$\sigma$ of solar composition whereas the CO is super-solar to $\sim$2$\sigma$. We additionally retrieve CO$_2$ and HCN but we see no constraints on either species given their weak opacity in the observed wavelength range.

The retrieved temperature profile for the hybrid retrieval is similar to the low-resolution retrieval but $\sim$1.5$\sigma$ away from the high-resolution constraint. This is because the HST and Spitzer data are the most sensitive to changes in temperature, particularly variations in the deep atmosphere where the continuum of the emergent spectrum is set. Therefore the photosphere near P$\sim$0.1~bar is in very close alignment with the low-resolution observations. The high-resolution observations on the other hand are able to probe the line cores and are thus more sensitive to the region at the top of the atmosphere. Hence the temperature begins to deviate the most away from the low-resolution retrieval in the upper atmosphere. The d$K_\mathrm{p}$ and d$V_\mathrm{sys}$ are also tightly constrained near the expected value of 0, indicating that the high-resolution observations are able to detect the planetary signal. We have additionally retrieved the scale factor $\log(a)$ as is done in the simulated dataset by \citet{brogi2019}. This parameter was included as a way for the hybrid retrieval to weight the likelihoods from the low-resolution and high-resolution data and we find that this is within 2$\sigma$ of the expected value of 0.

Our work demonstrates the potential of hybrid retrievals of exoplanetary atmospheres. Using high-resolution data improves the spectral coverage allowing the identification of trace species and the constraining of relative abundances. The low-resolution data is key to constraining the P-T profile and absolute abundances as it is able to provide information on the spectral continuum. As such there is remarkable synergy between the two approaches as reflected in the insights obtained from the results of the joint approach presented here. We have shown the simultaneous hybrid retrieval method has great potential for obtaining improved constraints on chemical abundances such as H$_2$O and CO. In particular, the hybrid approach has the potential to break degeneracies seen in low-resolution retrievals, such as between CO and CO$_2$ present in the 4.5~$\mu$m Spitzer data \citep{gandhi2018, kreidberg2014}. Such improvements will help in determining planetary C/O ratios with potential implications for the understanding of planet formation and/or migration histories \citep{oberg2011, madhu2014_formation, piso2016}. In the current work, we nominally constrain the C/O ratio to $0.991^{+0.009}_{-0.022}$ by determining the CO to CO+H$_2$O ratio, consistent with that seen in previous work \citep{brogi2017, brogi2019}.

Going forward, the potential for simultaneous hybrid retrievals are numerous. Firstly, the current TESS mission will provide many exoplanet candidates orbiting bright enough stars for characterisation with this method. Secondly, a wealth of high-resolution observations is expected with the commissioning of CRIRES+ at the VLT \citep{follert2014} as well as other high throughput instruments such as CARMENES \citep{quirrenbach2014}. The use of smaller telescopes with high throughput spectrographs such as GIANO has been shown to prove fruitful in HRS exoplanet spectroscopy \citep{brogi2018, guilluy2019}. Hybrid retrieval methods exploiting synergies between high and low-resolution data will be vital in fully exploiting the high-resolution spectra obtained with these facilities. The telescopes planned for the next decade will also require such a framework to fully exploit the observations they make. The prospect of JWST with increased sensitivity, spectral coverage and resolution will greatly improve the low-resolution constraints that are possible. From the ground the planned ELTs with vast collecting areas and equipped with high-resolution spectrographs can enhance the insights from JWST data through such hybrid retrieval methods. In the longer term, developing these methods to push the limits of exoplanet characterisation through such a holistic approach could enable the next generation of facilities to probe smaller potentially habitable worlds. 

\acknowledgments
SG, GH, and AP acknowledge support from the UK Science and Technology Facilities Council (STFC). We thank the anonymous referee for their helpful comments on our manuscript. This work makes use of observations made using the CRIRES spectrograph on the European Southern Observatory (ESO) Very Large Telescope (VLT) (186.C-0289). We thank the ESO Science Archive for providing the data.

\vspace{5mm}

\bibliography{ms}
\bibliographystyle{yahapj}
\end{document}